\documentclass[aps,prl,showpacs,twocolumn,superscriptaddress,longbibliography]{revtex4-1}
\usepackage[pdftex]{graphicx}
\usepackage[pdftex,bookmarks,colorlinks,breaklinks]{hyperref} 
\hypersetup{linkcolor=red,citecolor=blue,filecolor=dullmagenta,urlcolor=blue} 

\graphicspath{./figs/}
\DeclareGraphicsExtensions{.pdf,.jpeg,.png}
\newcommand{\eqref}[1]{eq.~(\ref{#1})}

\newcommand{\figref}[1]{fig.~\ref{#1}}
\newcommand{\Figref}[1]{Figure~\ref{#1}}
\newcommand{\llbracket}{\left\langle}
\newcommand{\rrbracket}{\right\rangle}
\begin{document}
\title{Wave Front Shaping in Quasi-One-Dimensional Waveguides}
\author{Julian B\"{o}hm}
\affiliation{Universit\'{e} Nice-Sophia Antipolis, CNRS, Laboratoire de Physique de la Mati\`{e}re Condens\'{e}e, UMR 7336, Parc Valrose, 06100 Nice, France}
\author{Ulrich Kuhl}
\email{ulrich.kuhl@unice.fr}
\affiliation{Universit\'{e} Nice-Sophia Antipolis, CNRS, Laboratoire de Physique de la Mati\`{e}re Condens\'{e}e, UMR 7336, Parc Valrose, 06100 Nice, France}

\begin{abstract}
Using 10 monopole antennas reaching into a rectangular multi mode waveguide we shape the incident wave to create specific transport even after scattering events. Each antenna is attached to an IQ-Modulator, which allows the adjustment of the amplitude and phase in a broad band range of 6-18 GHz. All of them are placed in the near field of the other, thus the excitation of an individual antenna is influenced by the presence of the other antennas. Still these 10 antennas are sufficient to generate any combination of the 10 propagating modes in the far field. At the output the propagating modes are extracted using a movable monopole antenna that is scanning the field. If the modes are scattered in a scattering region, the incident wave can be adjusted in such a way, that the outgoing wave can still be adjusted as long as localization is not present.
\end{abstract}

\maketitle

\section{Introduction}
Wave front shaping is now widely used in several fields of wave systems ranging from multi antenna Wifi communication using spatial multiplexing and spatial diversity to increase throughput~\cite{yu03}, to enhance visibility through tissues~\cite{vel08,vel10}, but also to destroy tumors by time reversal techniques~\cite{aub03}. This might be done in the temporal, spatial or combined sense. In this paper we will concentrate on the spatial part. Using wave front shaping one can create 'particle like scattering states'~\cite{rot11}, i.e. a diffusive field that has high intensity along a classical trajectory even though far away from a ray description like in classical optics. Other application might be in the field of coherent perfect absorbers~\cite{cho10,gma10} (CPA), also called anti-laser, and enhanced absorbers~\cite{wan11b} (CEA). These absorbers uses destructive interference as well as internal absorption to increase absorption. It can also used to generate full transmission through disordered~\cite{vel08,pen08,ger14} or enhance transmission even broad band~\cite{hsu15}. Using broad band wave front shaping one can focus waves even below subwavelength, which has been detailed in the frame of time reversal techniques and measured using ultrasound or microwaves~\cite{der01,ler06,ler07,arXdup16}.
In this paper we show that wave front shaping can be performed broad band in a rectangular waveguide using a monopole antenna array, which are fed by a vector network analyzer via IQ-Modulators, even if the antennas are placed in the near field. If the channel contains a scattering region wave front shaping is still possible.
However, depending on the scattering system and the outgoing wave it might not be broad band anymore.

\section{Experimental setup}
\begin{figure}
\centering \includegraphics[width=.99\linewidth]{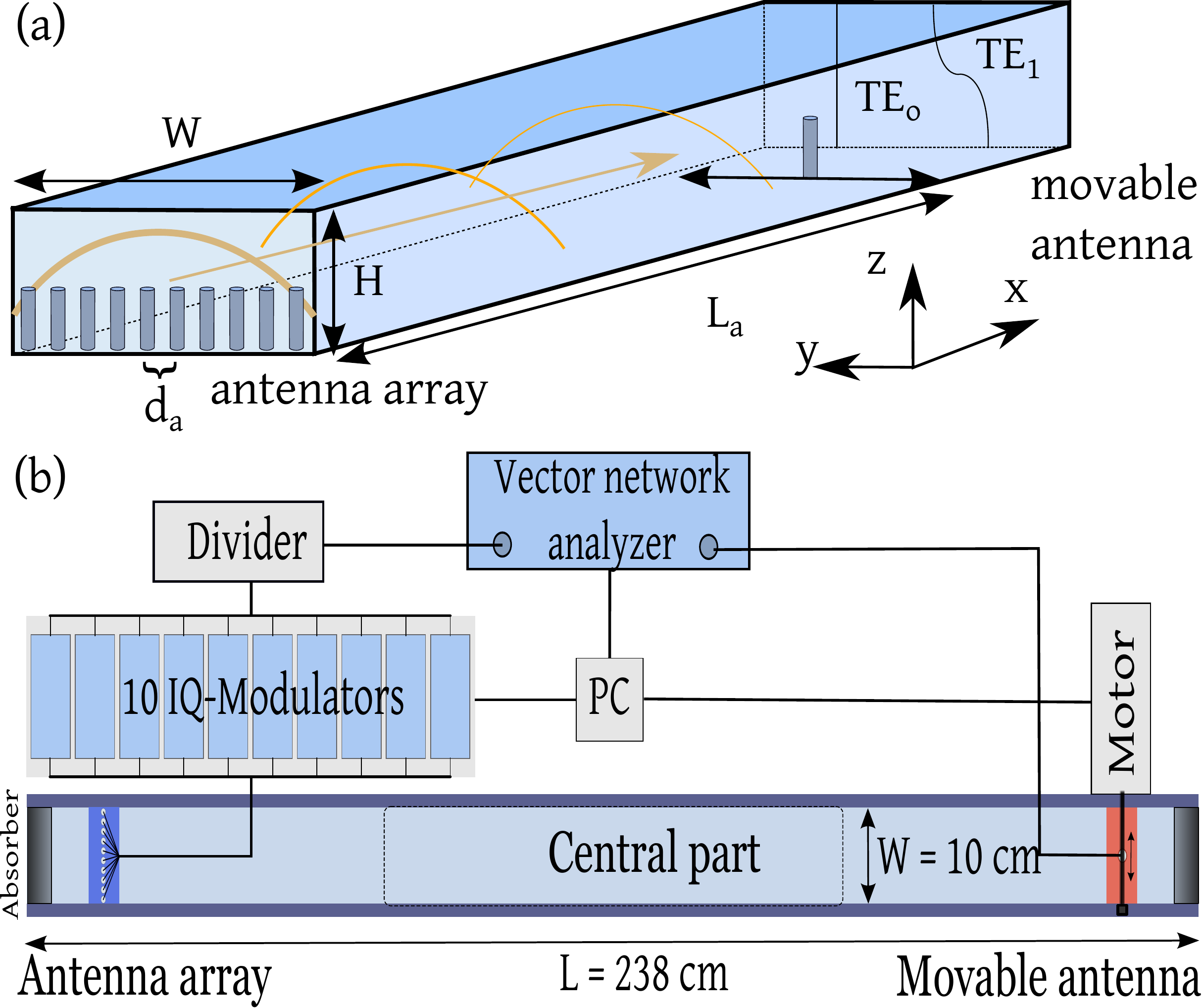}
\caption{\label{fig:SetUpSketch}
(a) Sketch of the waveguide indicating the antenna array and the movable antenna in a 3D sketch of a rectangular waveguide. The electric field dependence of the TE$_0$ and TE$_1$ modes are plotted as well. The height $H$=8\,mm. The antennas are monopole antennas introduced via small holes into the waveguide.
(b) The experimental setup including the PC, Vector Network Analyzer (VNA), cables, Power divider, IQ-Modulators, and the Q1D waveguide.
}
\end{figure}

\begin{figure*}
\centering \includegraphics[width=.99\linewidth]{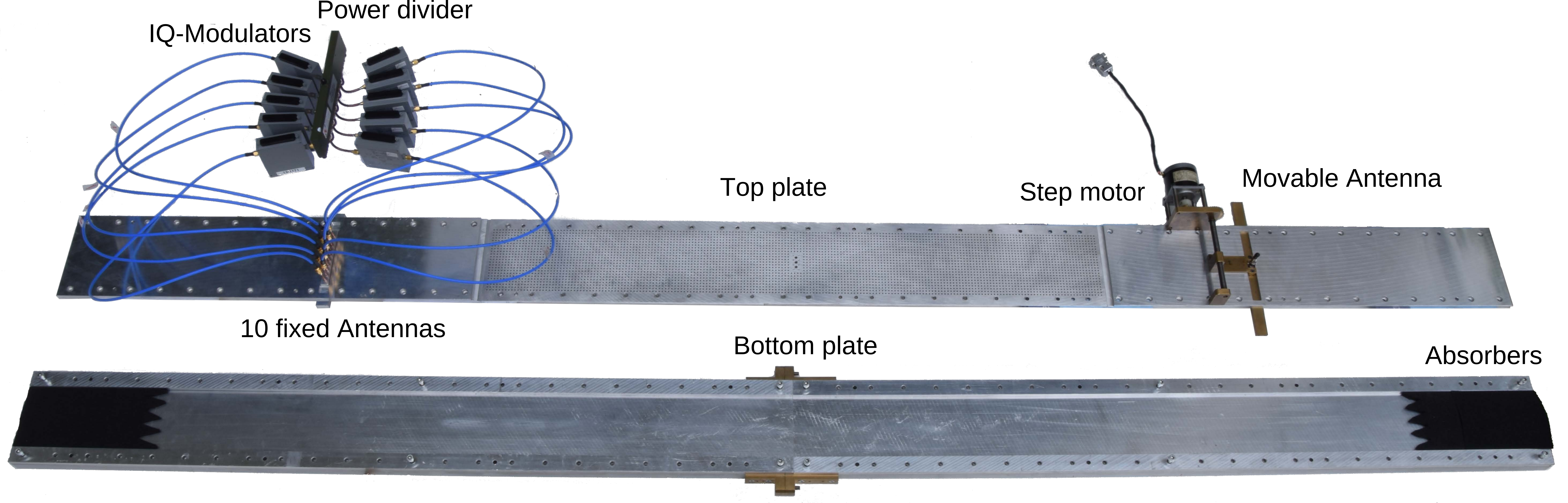}
\caption{\label{fig:photo}
A photograph of the experimental setup showing the exciting 10 antennas with their connection to the power divider via the IQ-Modulators. The top plate is moved from the bottom part to give a view of the interior of the waveguide showing the absorbers at both ends. The holes in the central region of the top plate make it possible to scan the field inside this region.
}
\end{figure*}

In this study we use an experimental setup that allows us to measure microwave transport through a multimode metallic-wall waveguide (see \figref{fig:SetUpSketch}(a)), where scatterers of different type can be placed. The waveguide has a height of $H=8$\,mm, a width $W=10$\,cm, a total length $L=2.38$\,m and a length between the antenna $L_a=1.50$\,m thus meeting the waveguide condition
\begin{equation}\label{eq:WaveguideDef}
  H\ll W \ll L.
\end{equation}
The quasi one-dimensional setup has been used to investigate the propagation of modes in correlated disordered systems~\cite{die11a,die12a,die12b}, using a second movable antenna instead of the antenna array.
In an empty rectangular waveguide the electromagnetic field can be separated into two components of different polarization:
the TM polarization with a transverse magnetic field and the TE polarization with a transverse electric field~\cite{jac62}.
To ensure only a single mode in $z$ direction all measurements are performed below the cut-off frequency $\nu_\textrm{cut}=c/2H\approx18.75$\,GHz.
In this case only the lowest TE-component, having the electric field stretched along the $z$ axis and the magnetic field being in the $(xy)$ plane (see the coordinate system in \figref{fig:SetUpSketch}), can propagate.

The input (transmitting) antenna array is excited by a vector network analyzer (Agilent E5071C) via a power splitter (Microot MPD16-060180) and IQ~vector-modulators (GTM~1~M2L-68A-5 of GT Microwave Inc). The IQ-modulators can be controlled by a PC and vary the phase and amplitude in the frequency range of 6-18\,GHz.
The field transmitted by each individual antenna can be adjusted by the corresponding IQ-Modulator, thus the ensemble excites a combination of propagating modes (see also photograph shown in \figref{fig:photo}).
It is worth noting that phase and amplitude control via the IQ-Modulators is broad band meaning that inducing an additional phase shift of $\Delta\phi$ and a relative amplitude reduction of $a$ is valid over the full operating range of the IQ-modulator.
$\Delta\phi$ and $a$ have a 12bit control and broad band stability of $\pm 2^\circ$ and $\pm 1$\,dB.

The output (receiving) monopole antenna is plugged into the waveguide via a slide that can be shifted stepwise by a motor (see \figref{fig:photo}).
Absorbers are placed on both ends of the waveguide in order to reduce reflections from the ends.
The used absorbers are ECCOSORB LS-14 and LS-16 with a relative impedance of 0.89 and 0.87 at 10 GHz, respectively.
The LS-14 is placed first in direction of the system and has a sawtooth form with 5 teeth in the $y$-direction and a sawtooth depth of about 3.5\,cm.
Its length is 8\,cm and thereafter the second absorber is placed using a linear interface and has a length of 14\,cm (for details see \figref{fig:photo}).
Both absorber fill the full height over the full width.
It is important to note that the absorbers at the end are placed sufficiently far away from the antennas to suppress evanescent coupling between antenna and absorbers.
In our case the distance is about 2.5 times the width $W$.
In the central region additional dielectric or metallic scatterer can be placed to include a scattering system.

\section{Individual Antenna Excitation}

Let us assume for the moment that within the antenna array the other antennas can be neglected.
Thus we treat the input antenna 1 as a point source, while the output antenna 2 is an observation point.
We will follow the description and notation of~\cite{die10a}.
The electric field of a point source is determined by the retarded Green's function ${\cal G}(|x-x'|;y,y')$.
The Green's function for the empty waveguide in the normal-mode representation reads
\begin{eqnarray}\label{eq:Gmode}
{\cal G}(|x-x'|;y,y')&=&
\sum_{n=1}^{N_W}A_n\sin\left(\frac{\pi n y}{W}\right) \sin\left(\frac{\pi n y'}{W}\right)\nonumber\\[6pt]
&&\times\frac{\exp\left(ik_n|x-x'|\right)}{ik_n W}.
\end{eqnarray}
Here we introduced the factor $A_n$ which effectively describes the coupling between the antennas and normal modes labeled by index $n$.
The quantities $\pi n/W$ and $k_n$ are, respectively, the discrete values of transverse, $k_y$, and longitudinal, $k_x$, wave numbers:
\begin{eqnarray}\label{eq:kn}
  k_n&=&\sqrt{k^2-\left(\frac{\pi n}{W}\right)^2}=\frac{2\pi}{c}\sqrt{\nu^2-(n\nu_c)^2}, \\\nonumber
  && \textrm{with}\ n=1,2,3,\ldots,N_W.
\end{eqnarray}
Here $k=2\pi\nu/c$ is the total wave number for the electromagnetic wave of frequency $\nu$.
The total number $N_W$ of propagating waveguide modes is determined by the integer part $\llbracket\ldots\rrbracket$ of the mode parameter
$kw/\pi=\nu/\nu_c$,
\begin{equation}\label{eq:Nw}
  N_W=\llbracket kW/\pi\rrbracket=\llbracket\nu/\nu_c\rrbracket.
\end{equation}
Correspondingly, the sum in \eqref{eq:Gmode} runs only over the propagating modes with $n\leq N_W$
and ignores the contribution of evanescent modes for which $n>N_W$. Indeed, for
evanescent modes the values of $k_n$ are purely imaginary; therefore, they do not
contribute to the transport. The critical frequency for the $n$th mode
\begin{equation}\label{eq:cutoff}
  \nu^{(n)}_c=nc/2W,
\end{equation}
below it is evanescent.

\begin{figure}
\centering \includegraphics[width=.99\linewidth]{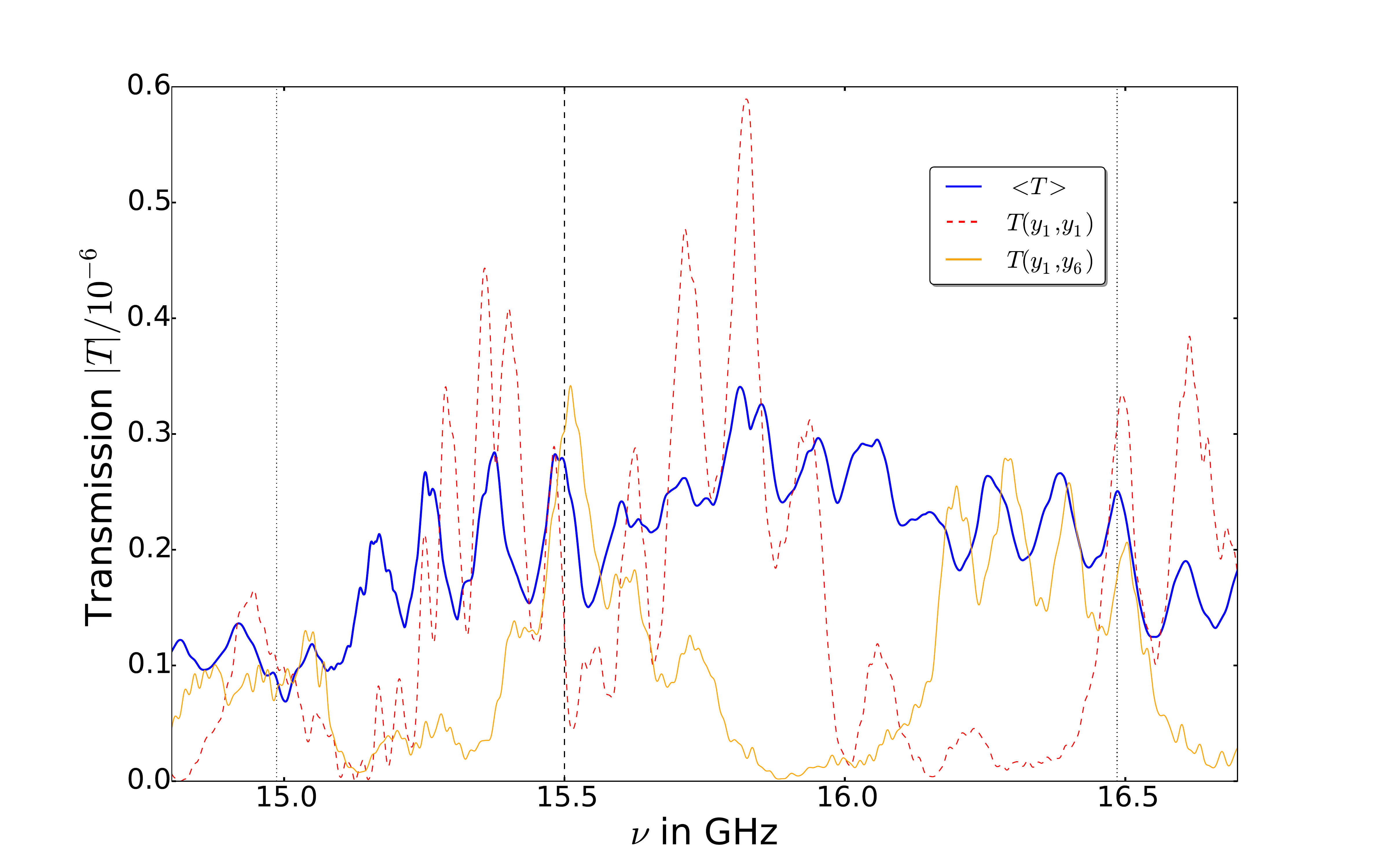}\\
\caption{\label{fig:TransmissionVsFrequency}
Transmission for different antenna positions $y_1$ and $y_2$ and its mean value $\langle T\rangle$ as a function of frequency, where the average was taken over all positions $y_1$ and $y_2$.
}
\end{figure}

\begin{figure}
\includegraphics[width=1.05\linewidth]{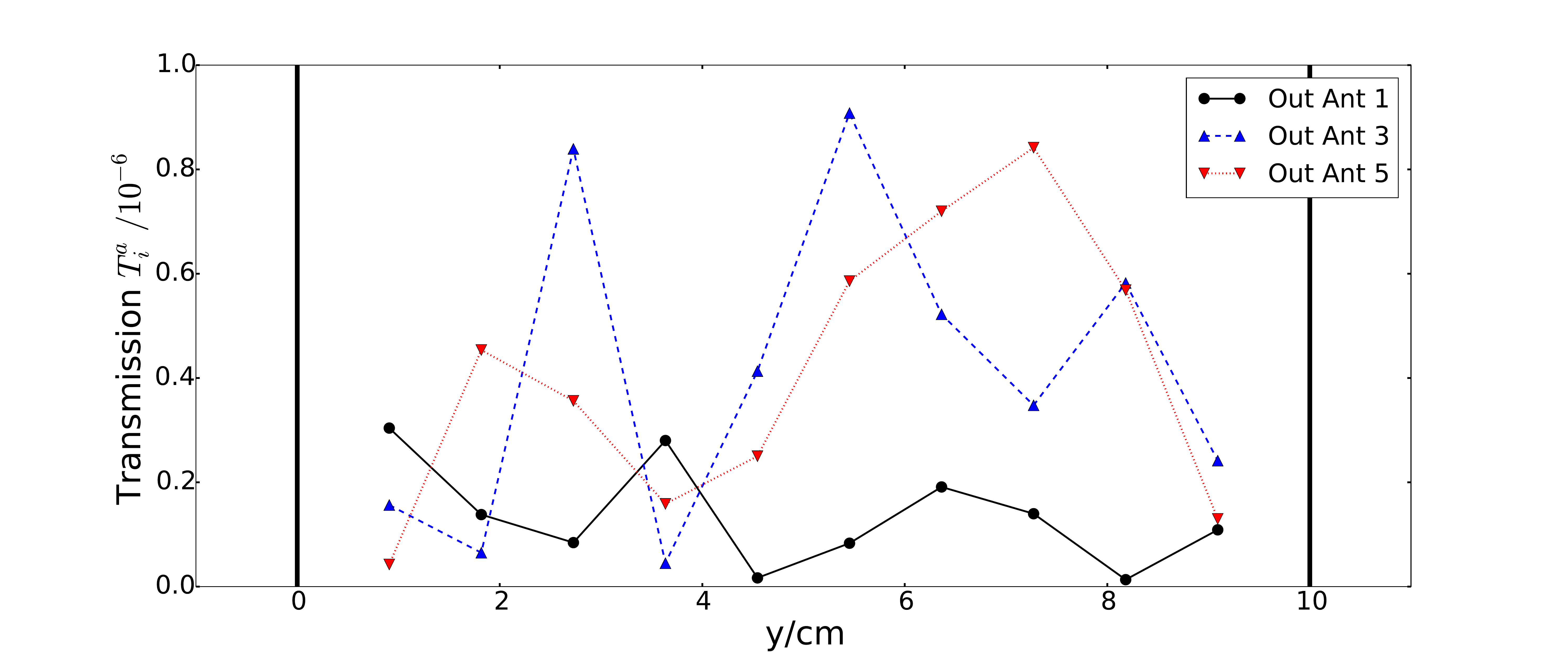}\\[-3ex]
\raisebox{3.cm}[0cm][0cm]{(a)}\\[0ex]
\includegraphics[width=1.05\linewidth]{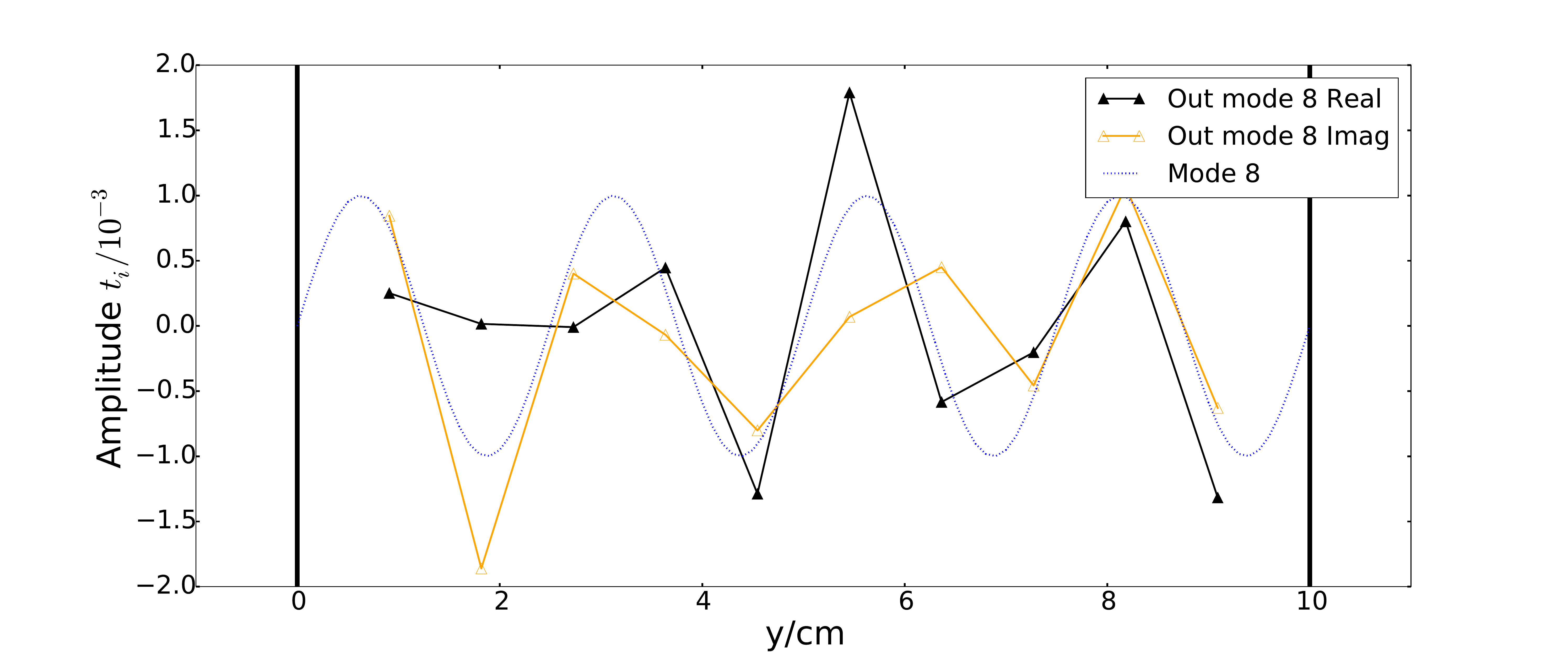}\\[-3ex]
\raisebox{3.cm}[0cm][0cm]{(b)}\\[0ex]
\caption{\label{fig:SingleAntExcitation}
(a) Measured transmission intensity $T^a_{ij}$ from different input antennas $i$ of the antenna array to the movable antenna at position $y_j$ at 15.5\,GHz.
(b) Real and imaginary parts of the transmission amplitudes $S_{nm}$ of the 8th mode (using \eqref{eq:Snm-def}) to the movable antenna at 15.5\,GHz. The dashed line corresponds to the 8th mode. A global phase shift to the data was performed such that the averaged imaginary part is 0.
}
\end{figure}

In this paper we will concentrate on the range $15\le\nu/$GHz$\le16.5$ where 10 modes can freely propagate.
The transmission through the rectangular waveguide for two different antenna positions and its averaged value is shown in Fig.~\ref{fig:TransmissionVsFrequency}.
The wave front shaping is adjusted at 15.5\,GHz.

The scattering matrix $S$ in the modal representation for the whole waveguide ($|x-x'|=L$) can be described as the twofold sine-Fourier transform,
\begin{equation}\label{eq:Snm-def}
  S_{nm}=\frac{2}{W}\int\limits_0^W \textrm{d}y\textrm{d}y' \sin\left(\frac{\pi n y}{W}\right) \sin\left(\frac{\pi m y'}{W}\right) {\cal G}(L;y,y').
\end{equation}
One can see that in accordance with \eqref{eq:Gmode} the scattering matrix for the empty quasi-1D waveguide \eqref{eq:Snm-def} is diagonal in the mode representation.

For the experimental data analysis, the continuous Fourier transform \eqref{eq:Snm-def} is replaced by its discrete counterpart, allowing us to compute
$S_{nm}$ from ${\cal G}(L;y,y')$ measured for different positions $y$ and $y'$ of input and output antennas.
An example of the transmission at 15.5\,GHz is shown in \figref{fig:SingleAntExcitation}.
In (a) the transmission intensity of a single antenna excitation of the input array is shown as a function of the output antenna position for three different
input antennas. In (b) we use \eqref{eq:Snm-def} to excite only the 8th mode of the cavity. As in the rectangular waveguide the 8th mode is freely propagating we expect to see the 8th mode only. Clearly we do not observe only the 8 mode, but a superposition of several modes. The relative proportion of the 8th mode is 64\% of the total intensity. Have in mind that using the same kind of scanning antenna \eqref{eq:Snm-def} is valid and have been used to investigate correlated disorder in quasi one-dimensional systems~\cite{die11a,die12a,die12b}.
This effect is due to the presence of the other antennas in the near field, which destroys the simple description \eqref{eq:Snm-def}.
In the next section, we will show experimentally that it is still possible to excite all 10 modes individually even for the whole frequency range of 15 to 16.5\,GHz, referring exactly to 10 propagating modes.

\section{Decomposition in Rectangular Waveguide and Broad Band Characteristics}

\begin{figure}
\centering \includegraphics[width=.95\linewidth,height=.65\linewidth]{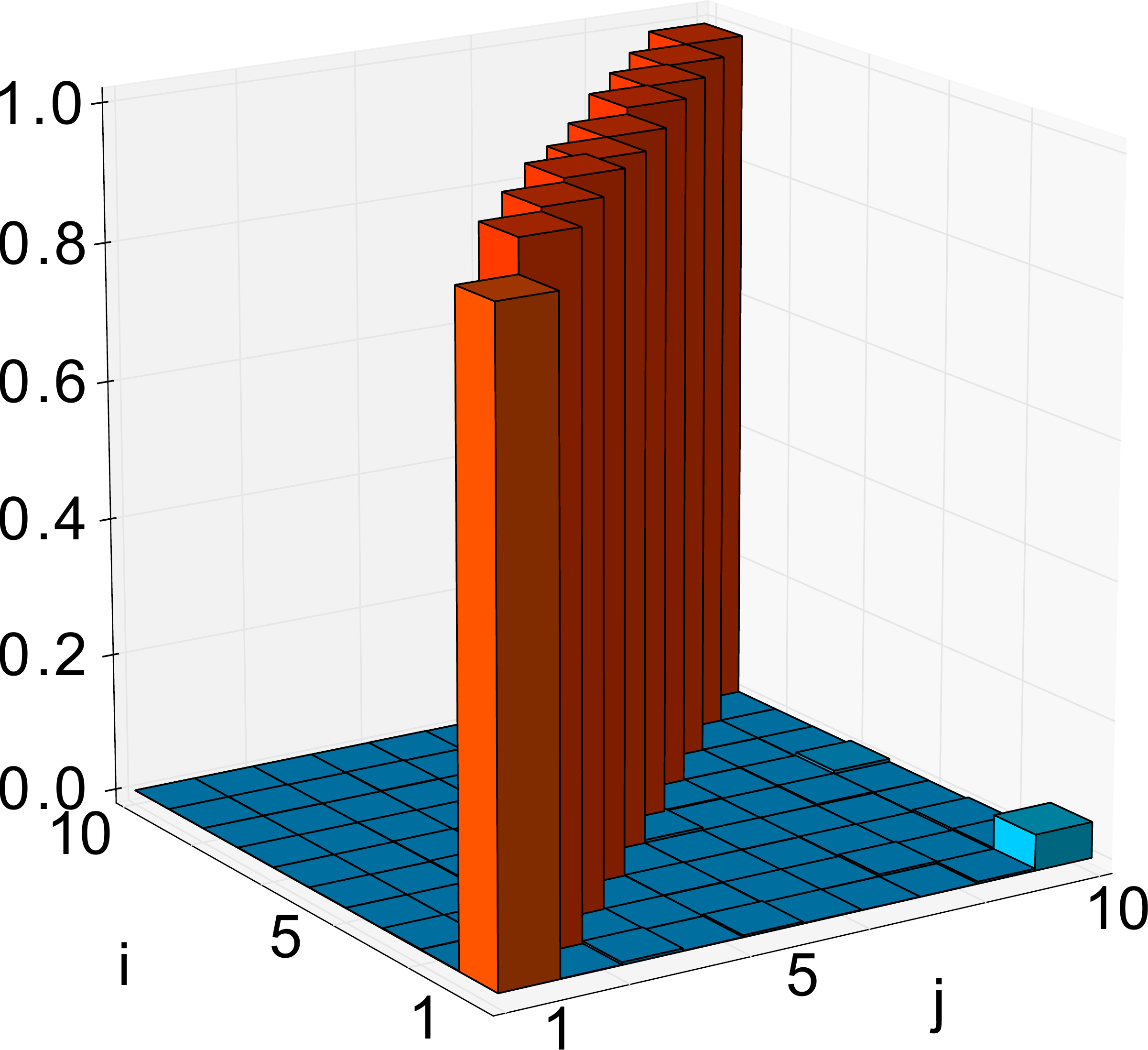}
\caption{\label{fig:RectTMatrix}
Measured Transmission matrix elements $T_{ij}$ for the rectangular waveguide where $i$ represent the sinusoidal mode number of the input and j the sinusoidal mode number of the output. The input basis have been adjusted by appropriate IQ-modulator phases and amplitudes.
}
\end{figure}

We now measure the transmission of the antenna array to the movable antenna by illuminating each antenna individually.
Each excitation of the other antennas are suppressed by about 40\,dB using the IQ-modulators (see also \figref{fig:SingleAntExcitation}(a)).
By solving a linear equation system for sinusoidal output we calculated the new basis for the input antennas.
This calculation yields the phase $\phi_i$ and amplitudes $a_i$ values to adjust the IQ-modulators.
Now we are able to excite exactly the mode of interest.
In \figref{fig:RectTMatrix} the transmission matrix $T$ in the sinusoidal input and output basis is shown showing the diagonal matrix.

\begin{figure}
\includegraphics[width=1.05\linewidth]{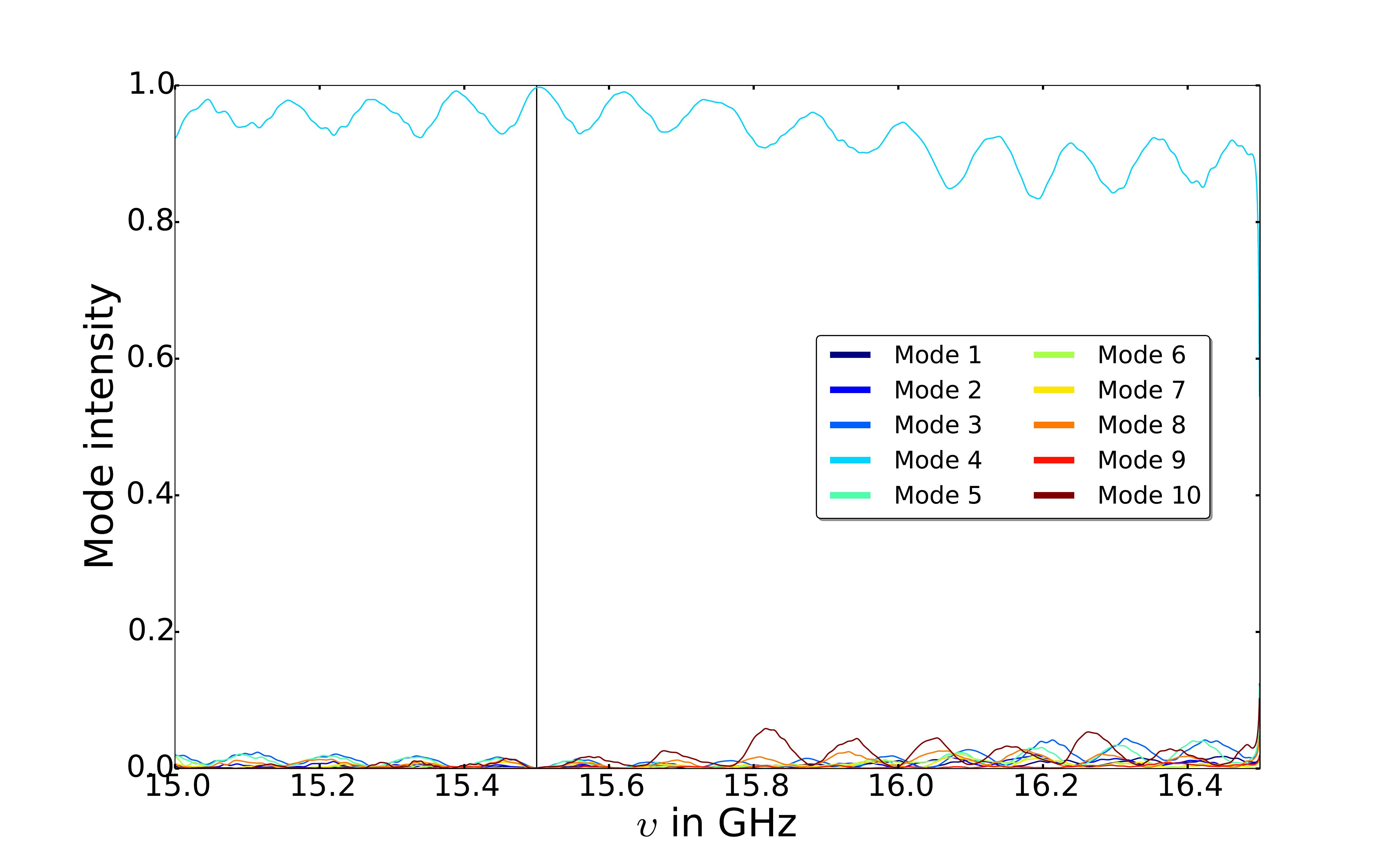}\\[-3ex]
\raisebox{4.5cm}[0cm][0cm]{(a)}\\[0ex]
\includegraphics[width=1.05\linewidth]{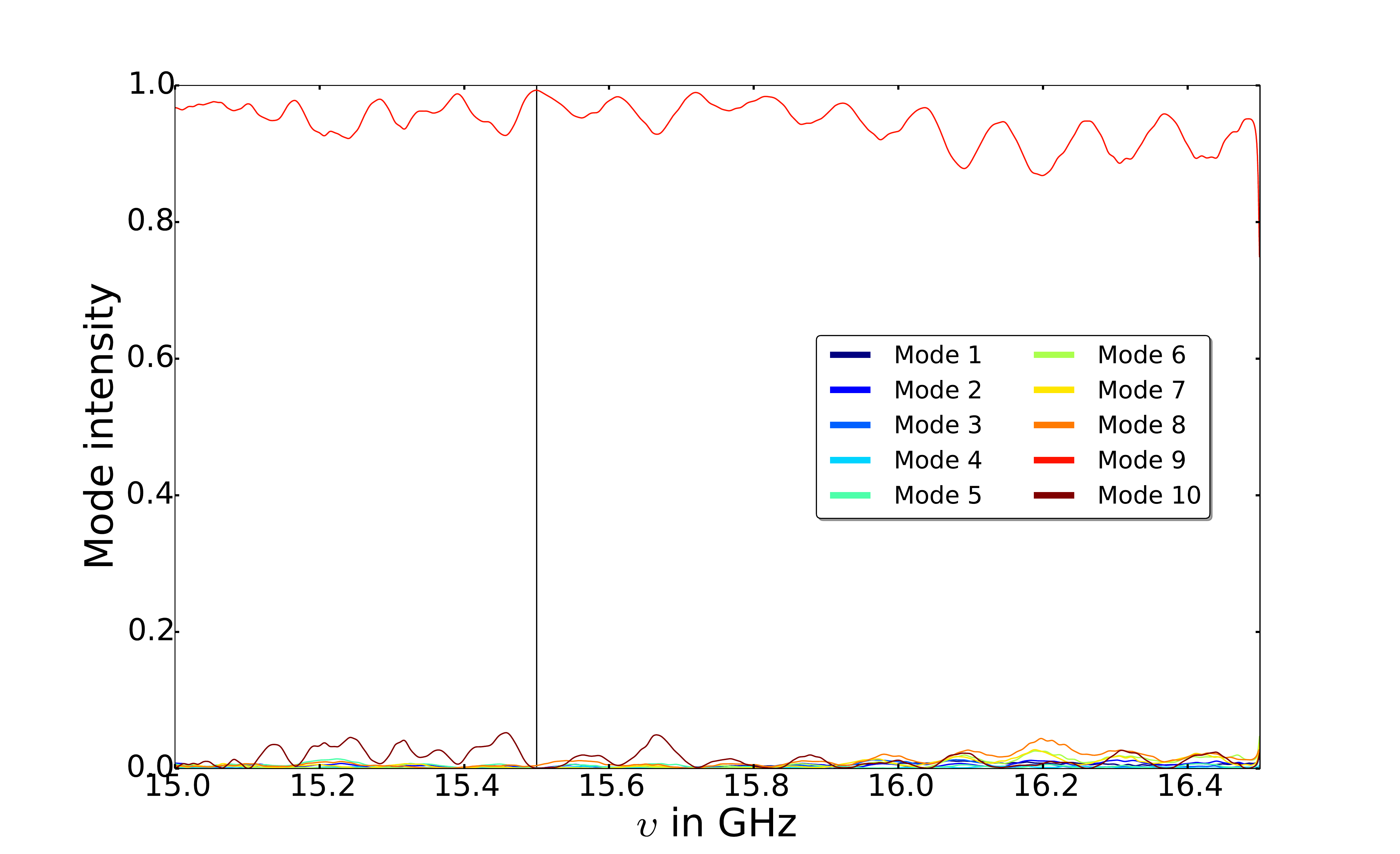}\\[-3ex]
\raisebox{4.5cm}[0cm][0cm]{(b)}
\caption{\label{fig:WvShapingRect}
Measured transmission $T=|t(n_\mathrm{in},n_\mathrm{out},\nu)|^2$ of incident mode $n_\mathrm{in}=3$ (a) and 8 (b) as a function of frequency $\nu$.
Over the whole range, where 10 modes are supported, mainly the outgoing mode $\mathrm{out}=3$ or 8, respectively, is dominant, even though the adjustment was done for $\nu=15.5$\,GHz.
}
\end{figure}

As the IQ-modulators adjust the phase and amplitudes in the whole frequency range from 6 to 18\,GHz and we have taken care of using the same cables, connectors, antennas, so the phase shifts for all antennas should be the same over a broad frequency range.
In \figref{fig:WvShapingRect} we present the transmission of input mode 3 (a) and 8 (b) into all modes over the frequency range supporting exactly the 10 modes.
On the one hand side one observes that at the adjustment frequency 15.5\,GHz the modes 3 and 8 are very close to 1, respectively.
Apart from that, mode 3 (8) are still very dominant in the whole range presented.

\section{Decomposition with Scattering region}
\begin{figure}
\centering \includegraphics[width=.95\linewidth,height=.65\linewidth]{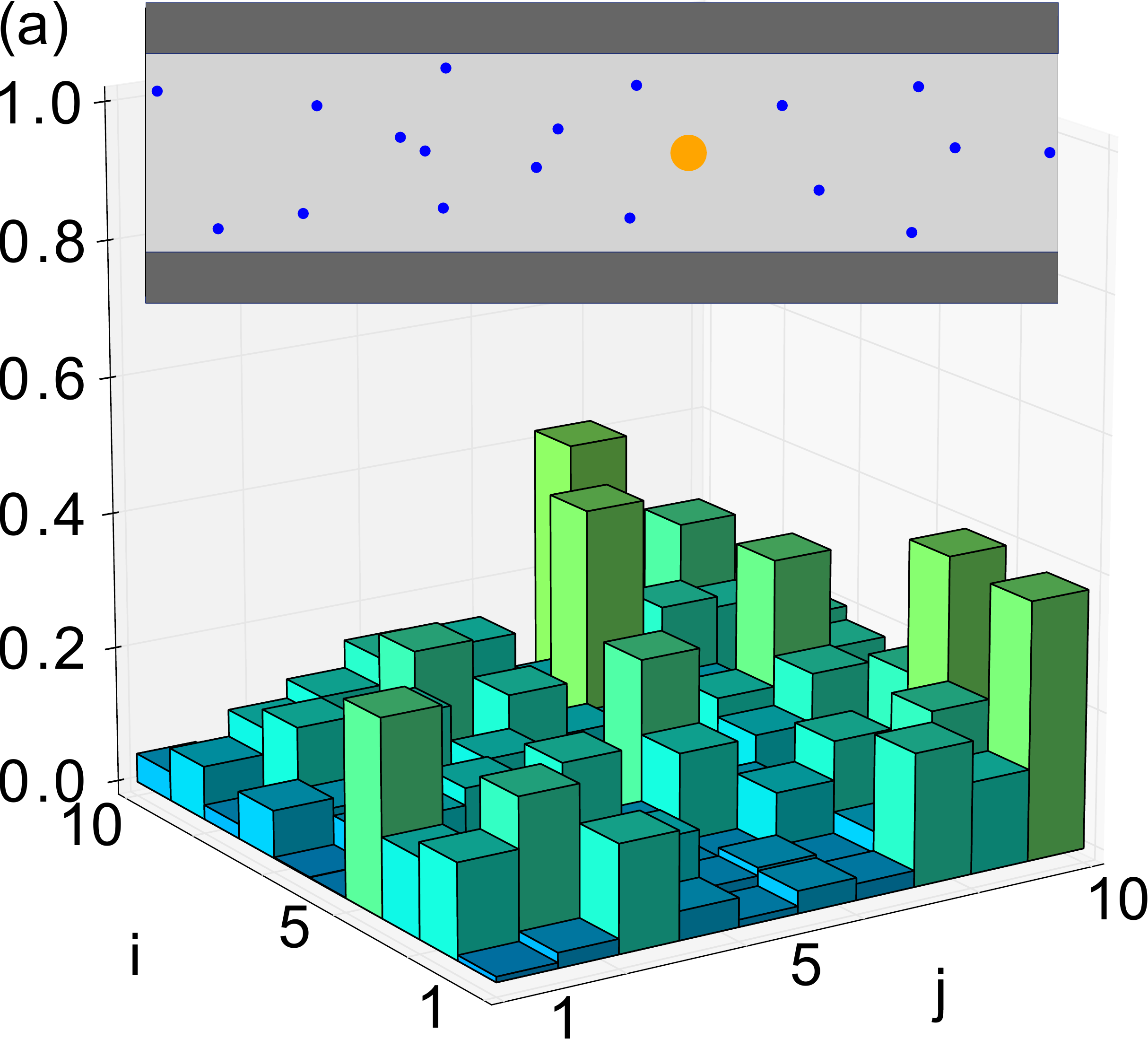}
\centering \includegraphics[width=.95\linewidth,height=.65\linewidth]{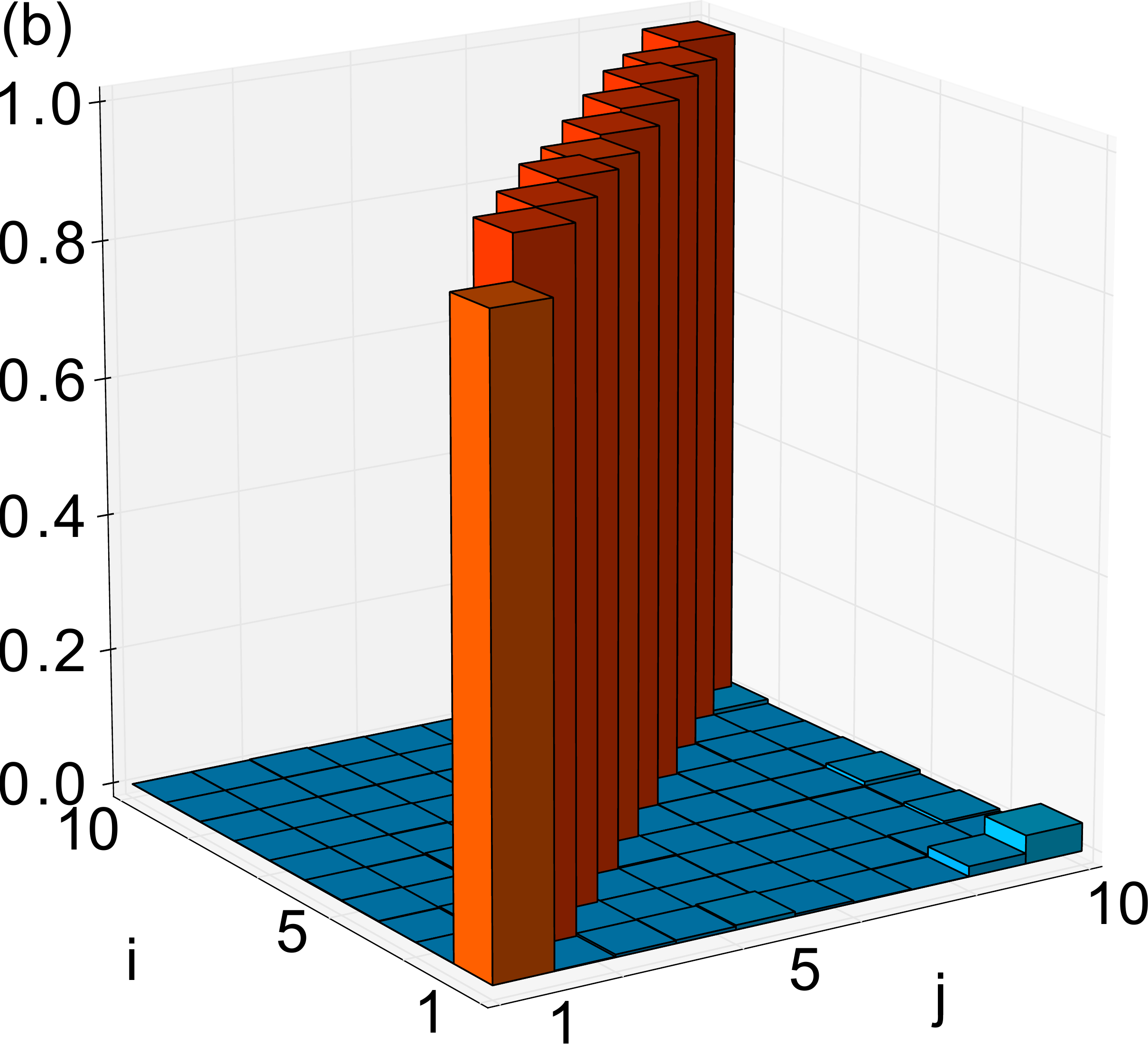}
\caption{\label{fig:TMatrixScatRegion}
Introducing a scattering region with 18 teflon cylinders with $r=2.5$\,mm (blue filled circles in the inset) and with a larger brass cylinder with radius $r=8.85$\,mm (orange filled circle in the inset) a single mode excitation $n_\mathrm{in}$ is scattered into many outgoing channels $n_\mathrm{out}$ (see upper part).
Readjusting the incoming wave front to a single outgoing mode $n_\mathrm{out}$ can be created.
}
\end{figure}

Now we include a scattering region inside the rectangular waveguide (see inset \figref{fig:TMatrixScatRegion}(a) and \figref{fig:SetUpSketch}(b)), where we place 18 dielectric (Teflon) scatterers with radius 2.5\,mm and a larger brass scatterer inside the central region.
\Figref{fig:TMatrixScatRegion}(a) shows the transmission matrix $T$ in the sinusoidal input and output basis.
The scatterers distribute the incoming modes into several outgoing modes. Using this scattering information one can readjust the incident basis in such a way that we get single sinusoidal modes on the outgoing part, leading to a diagonal transmission matrix $T'$, where the $'$ indicates the change of basis (see \Figref{fig:TMatrixScatRegion}(b)).

\begin{figure}
\includegraphics[width=1.05\linewidth]{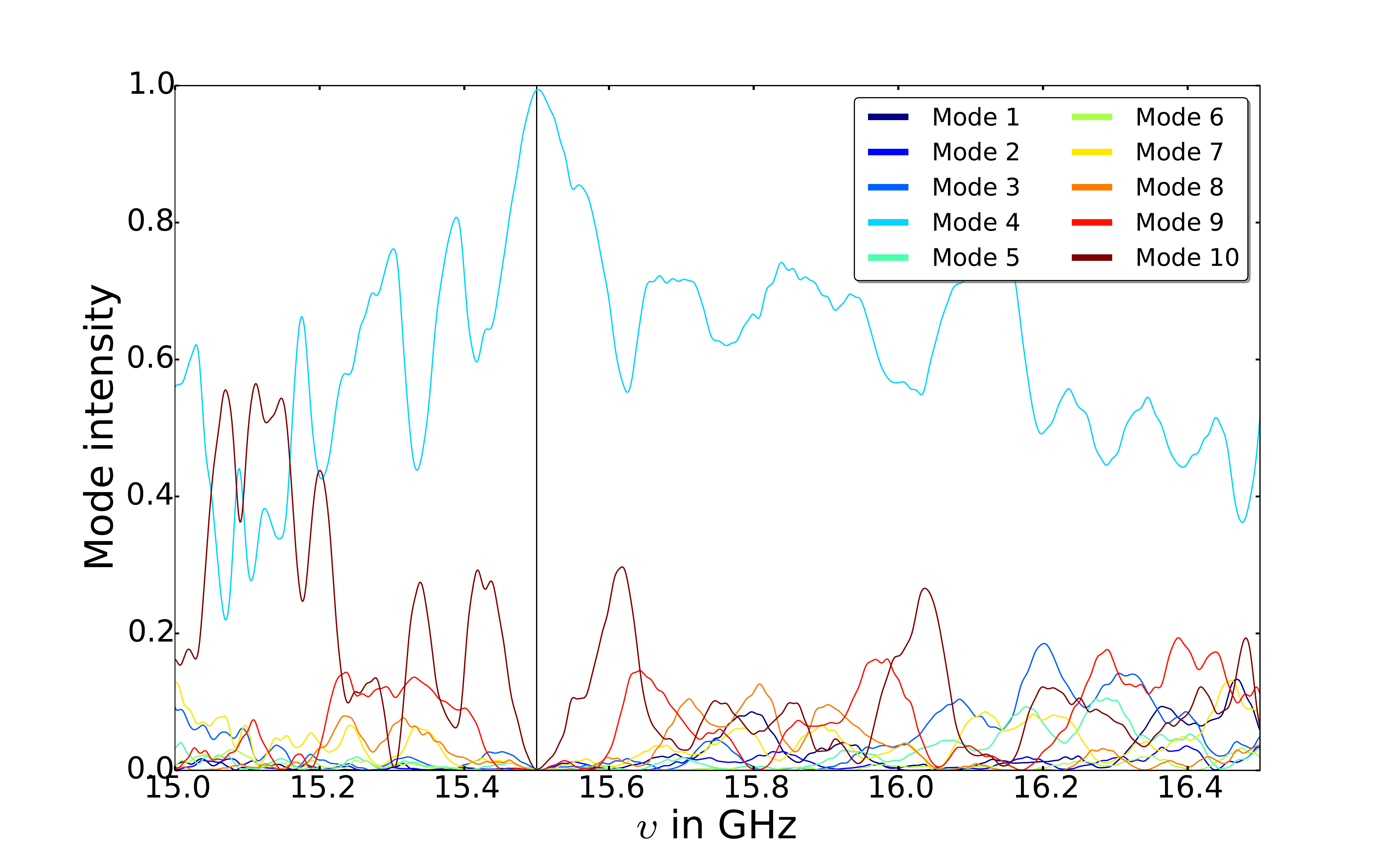}\\[-3ex]
\raisebox{4.5cm}[0cm][0cm]{(a)}\\[0ex]
\includegraphics[width=1.05\linewidth]{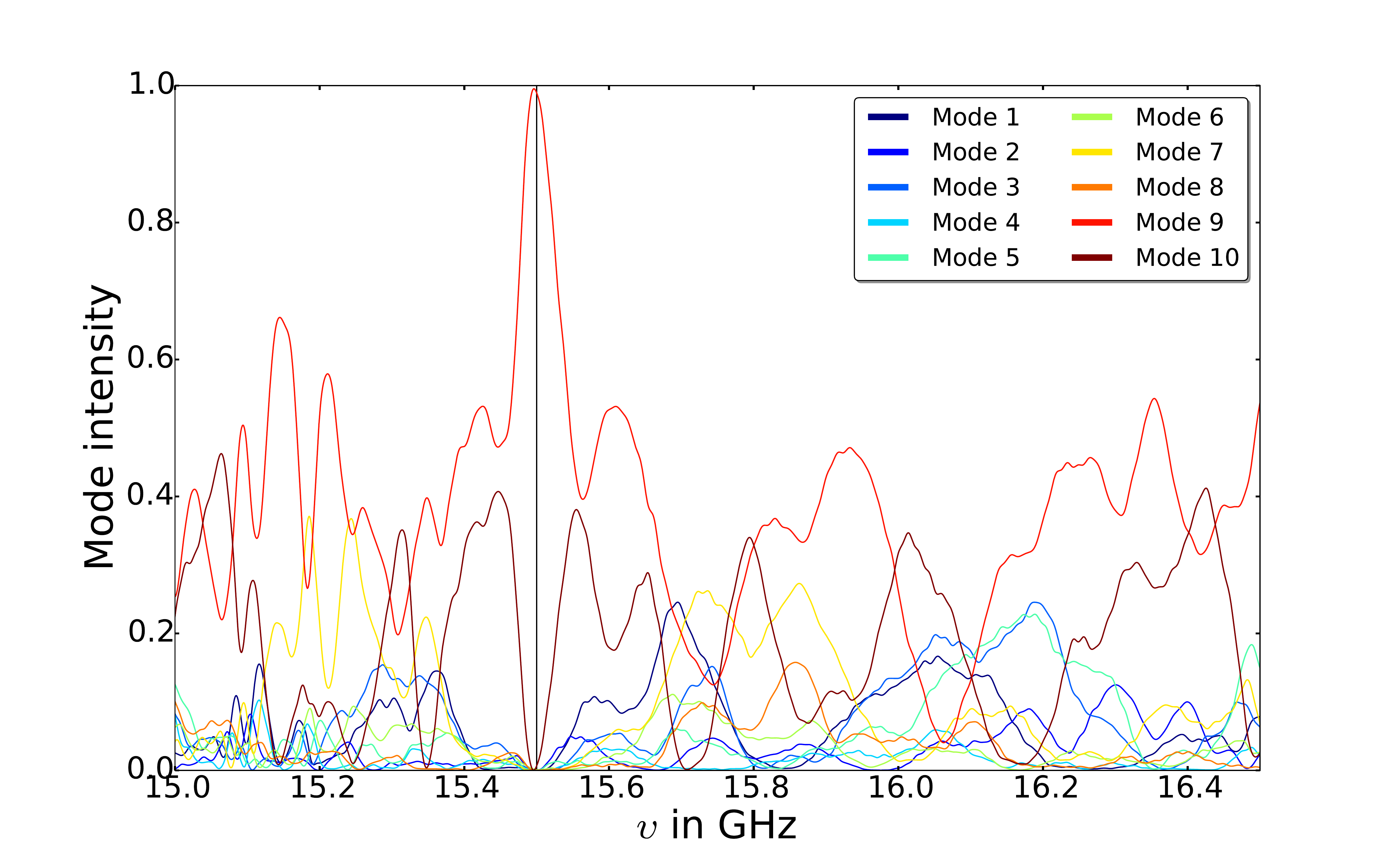}\\[-3ex]
\raisebox{4.5cm}[0cm][0cm]{(b)}\\[0ex]
\caption{\label{fig:WvShapingRect+Scat}
Measured transmission $T'=|t(n'_\mathrm{in},n_\mathrm{out},\nu)|^2$ of incident mode $n'_\mathrm{in}=3$ (a) and 8 (b) as a function of frequency $\nu$.
Over the whole range, where 10 modes are supported. The outgoing mode $\mathrm{out}=3$ or 8 are still dominant close to the adjustment frequency $\nu=15.5\,GHz$. But in case of the 8th mode it decays rapidly loosing the broad band features whereas mode 3 stays dominant over a broader frequency range.
}
\end{figure}

Depending on the scattering system certain outgoing modes might be still rather broad band. This can be seen in \figref{fig:WvShapingRect+Scat}, where the 8th mode is very narrow but the $n_\mathrm{out}=3$ mode is still broad, even though not comparable to the situation observed in the rectangular waveguide (see \figref{fig:WvShapingRect}).
Note that it is important that the scattering systems transmits sufficiently and localization effects are still weak.
If the scattering system has not sufficient transporting modes, i.e. the transporting modes are smaller then the number of outgoing modes, a full control will not be feasible.
In the case of strong localization the control of the outgoing mode by incident wave shaping is totally lost as the outgoing mode is solely defined by a single transporting mode~\cite{pen14}.

\section{Conclusion}
In this paper we have shown that it is possible to make broad band wave front shaping within a multi mode waveguide, e.g.\ a quasi one-dimensional rectangular waveguide, by using IQ-modulators and monopole antennas. The fact, that the antennas are situated in the near field of each other does destroy the use of sine transforms to adjust the IQ-modulators.
However, full potential of wave front shaping is accessible if the far field scattering matrix of each individual antenna is known by measurement.
We were even able to perform the incident wave shaping in such a way that the wave after undergoing scattering shows a pure sinusoidal mode profile.
Attaching the quasi one-dimensional channel including the antenna array with the IQ-modulators to different kind of systems gives the possibility to enhance transmission, realize particle like scattering states, create coherent perfect absorbers and focus microwaves on sub wave length scales as already indicated in the introduction.

\section*{Acknowledgment}

The authors would like to thank the ANR for funding via the ANR GEPARTWAVE Project (ANR-12-IS04-0004-01) and
the European Commission through the H2020 programme by the Open Future Emerging Technology "NEMF21" Project (664828).

\end{document}